\documentclass[11pt,a4paper,useAMS,usenatbib]{emulateapj}
\usepackage{amsmath}
\usepackage{epsfig}
\usepackage{natbib}
\usepackage{xcolor}
\usepackage{epsfig}
\usepackage{graphicx}
\usepackage{pifont}
\usepackage{xcolor}
\usepackage{lineno}
\usepackage{hyperref}

\begin{document}

\title{A candidate supermassive black hole in a gravitationally-lensed galaxy at $z\approx10$}

\author{Orsolya E. Kov\'acs\altaffilmark{1}, \'Akos Bogd\'an\altaffilmark{2}, Priyamvada Natarajan\altaffilmark{3,4,5}, Norbert Werner\altaffilmark{1}, \\ Mojegan Azadi\altaffilmark{2}, Marta Volonteri\altaffilmark{6}, Grant R. Tremblay\altaffilmark{2}, Urmila Chadayammuri\altaffilmark{2}, \\ William R. Forman\altaffilmark{2}, Christine Jones\altaffilmark{2}, Ralph P. Kraft\altaffilmark{2}}
\affil{\altaffilmark{1} Department of Theoretical Physics and Astrophysics, Faculty of Science, Masaryk University, Kotl\'a\v{r}sk\'a 2, Brno, 611 37, Czech Republic}
\affil{\altaffilmark{2} Center for Astrophysics \ding{120}  Harvard \& Smithsonian, 60 Garden Street, Cambridge, MA 02138, USA}
\affil{\altaffilmark{3} Department of Astronomy, Yale University, New Haven, CT 06511, USA}
\affil{\altaffilmark{4} Department of Physics, Yale University, New Haven, CT 06520, USA}
\affil{\altaffilmark{5} Black Hole Initiative, Harvard University, 20 Garden Street, Cambridge, MA 02138, USA}
\affil{\altaffilmark{6} Institut d'Astrophysique de Paris, Sorbonne Universit\'e, CNRS, UMR 7095, 98 bis bd Arago, 75014 Paris, France}

\shorttitle{SUPERMASSIVE BLACK HOLE IN A $z\approx10$ GALAXY}
\shortauthors{KOV\'ACS ET AL.}

\begin{abstract}
While supermassive black holes (BHs) are widely observed in the nearby and distant universe, their origin remains debated with two viable formation scenarios with light and heavy seeds. In the light seeding model, the first BHs form from the collapse of massive stars with masses of $10-100 \ \rm{M_{\odot}}$, while the heavy seeding model posits the formation of $10^{4-5} \ \rm{M_{\odot}}$ seeds from direct collapse. The detection of BHs at redshifts $z\gtrsim10$, edging closer to their formation epoch, provides critical observational discrimination between these scenarios. 
Here, we focus on the \textit{JWST}-detected galaxy, GHZ\,9, at $z\approx10$ that is lensed by the foreground cluster, Abell\,2744. Based on 2.1~Ms deep \textit{Chandra} observations, we detect a candidate X-ray AGN, which is spatially coincident with the high-redshift galaxy, GHZ\,9. The BH candidate is inferred to have a bolometric luminosity of $(1.0^{+0.5}_{-0.4})\times10^{46} \ \rm{erg \ s^{-1}}$, which corresponds to a BH mass of $(8.0^{+3.7}_{-3.2})\times10^7 \ \rm{M_{\odot}}$ assuming Eddington-limited accretion. This extreme mass at such an early cosmic epoch suggests the heavy seed origin for this BH candidate.
Based on the \textit{Chandra} and \textit{JWST} discoveries of extremely high-redshift quasars, we have constructed the first simple AGN luminosity function extending to $z\approx10$. Comparison of this luminosity function with theoretical models indicates an over-abundant $z\approx10$ BH population, consistent with a higher-than-expected seed formation efficiency.
\end{abstract}


\section{Introduction}
\label{sec:intro}

While BHs play a central role in the evolution of galaxies, their origin remains unsettled. Historically, two main formation scenarios can describe the birth of the first BHs: the light \citep[e.g.][]{2001ApJ...551L..27M,2003ApJ...582..559V} and the heavy seed channels \citep[e.g.][]{2005ApJ...633..624V,2006MNRAS.370..289B,2006MNRAS.371.1813L,2019Natur.566...85W}. In the former channel, seed BHs form from the collapse of population III stars with $10-100 \ \rm{M_{\odot}}$. Although seeds are abundant in this channel, growing them by $5-7$ orders of magnitude in mass is challenging and may require sustained super-Eddington accretion. 
As an alternative to the traditional light seed model, a population of $10^3-10^4\,\rm  M_{\odot}$ seeds forming in dense stellar structures was proposed \citep[e.g.][]{2020OJAp....3E..15R,2022MNRAS.512.6192S}.
Per the heavy seed channel, BHs could originate from the collapse of massive gas clouds and end up with birth masses of $10^4-10^5 \ \rm{M_\odot}$.

Several physical processes and scenarios have been invoked to explain the direct collapse of pristine gas in primordial halos. Early theoretical work suggested the irradiation of pristine atomic cooling halos by Lyman-Werner photons to suppress gas cooling and fragmentation \citep{lodato06} as prerequisites for direct collapse.
Recent simulations suggest that multiple feasible pathways exist to form heavy initial BH seeds from direct collapse: 
Cold flows that are expected to occur naturally in standard structure formation at early epochs could lead to direct collapse and BH seed formation \citep{Latif+2022}. The rapid amplification of light seeds in specific environments like gas-rich, dense nuclear star clusters can also result in the formation of heavy seeds within a few million years \citep{Tal+2014}.
\citet{2019Natur.566...85W} suggests that dynamical heating in rapidly growing pre-galactic gas clouds is the main driver to form seeds with masses of $\sim\!10^4 \, \rm M_{\odot}$.
Recent work by \citet{Mayer+2023} describes a very rare scenario, in which the merger of gas-rich galaxies leads to the formation of dense, compact supermassive disks that can also directly collapse into $10^{6-8}\,\rm{M_{\odot}}$ BHs.

To probe the origin of BH seeds, the detection of $z\gtrsim9$ BHs provides the cleanest observational test \citep{2018MNRAS.481.3278R}. Until recently, the high-redshift BH population has remained hidden. However, \textit{Chandra} and \textit{JWST} observations have started to reveal the presence of a handful of candidate AGN in the very early universe \citep{bogdan2023detection,2023ApJ...953L..29L,2023arXiv230512492M}.

\begin{figure*}[!htp]
  \begin{center}
    \leavevmode
      \epsfxsize=0.85\textwidth \epsfbox{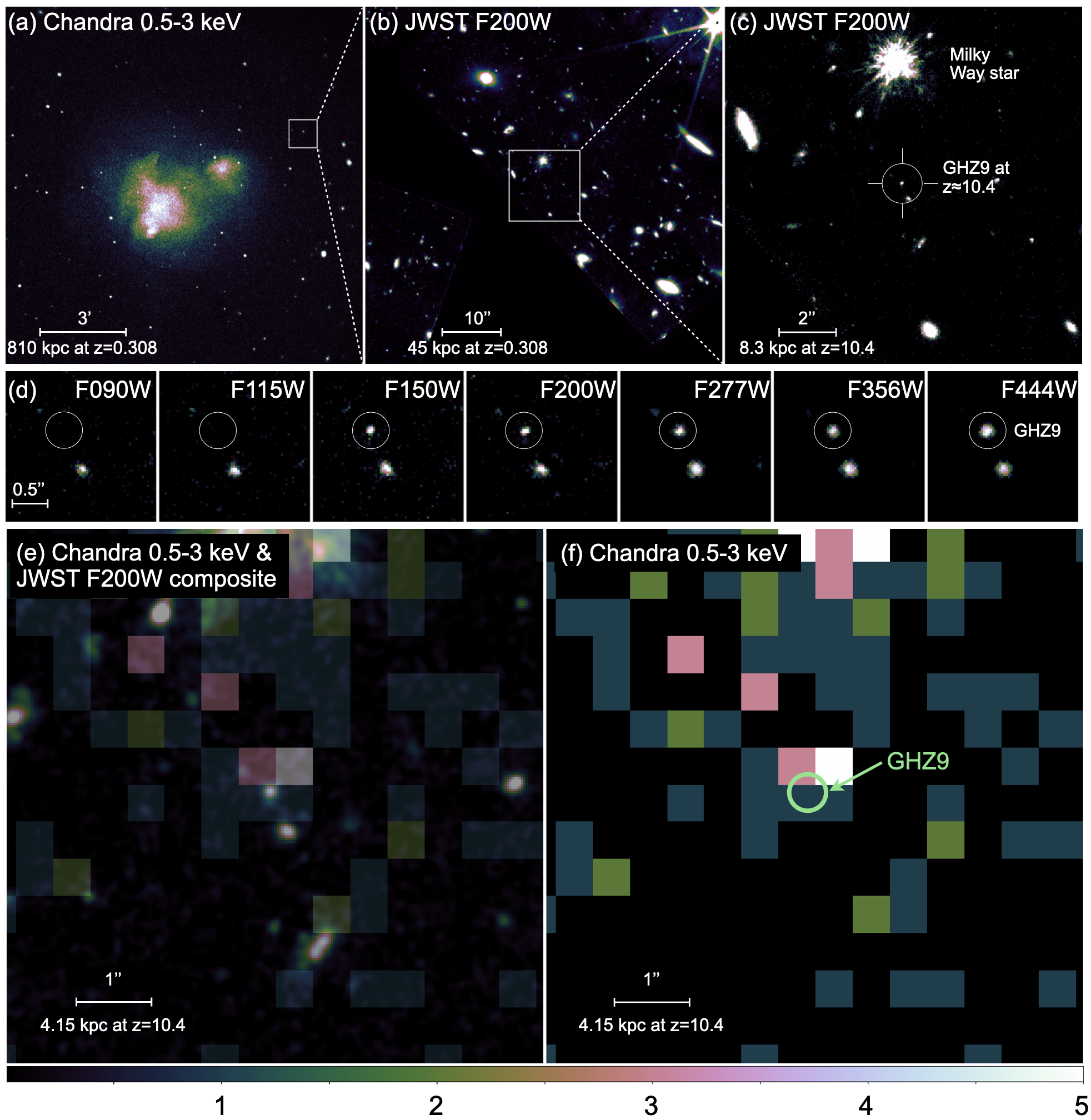}
      \vspace{0cm}
      \caption{\textit{Chandra} ACIS and \textit{JWST} NIRCam images of the $z \approx 10.4$ galaxy, GHZ\,9 and its surroundings. Panel (a) shows the $0.5-3$\,keV band X-ray mosaic of Abell\,2744 marking the position of GHZ\,9. Panels (b) and (c) are zoomed-in \textit{JWST} NIRCam images showing the surroundings of GHZ\,9. Panel (d) presents postage stamp NIRCam images of GHZ\,9 and its projected neighbor, a $z\approx 1.6$ galaxy, in seven bandpass filters. GHZ\,9 is not detected in the two shortest wavelengths, indicating its Lyman-break nature. Panel (e) is the \textit{Chandra} and \textit{JWST} composite image of the immediate vicinity of GHZ\,9. The $z\approx10.4$ galaxy, GHZ\,9, is co-spatial with the X-ray source, while panel (f) is the \textit{Chandra} counts image of GHZ\,9.} 
     \label{fig:ghz9_jwst}
  \end{center}
\end{figure*}

We probe high-redshift BHs by exploring faint X-ray sources that are magnified by gravitational lensing behind the lensing galaxy cluster, Abell\,2744 ($z=0.308$).
The \textit{JWST} UNCOVER (Ultradeep NIRSpec and NIRCam ObserVations before the Epoch of Reionization) survey reported the detection of $19$ high-redshift ($z>9$) lensed galaxies behind Abell\,2744 \citep{atek23}. Using deep \textit{Chandra} X-ray observations, we follow up these high-redshift galaxies and search for X-ray emission associated with their central BHs.

We search for X-ray emission co-spatial with the 19 $z>9$ lensed galaxy candidates in the $0.5-3$~keV (soft), $3-7$~keV (hard), and $0.5-7$~keV (broad) band merged \textit{Chandra} images.
We emphasize that this is a preliminary search designed solely to identify possible X-ray sources.
We identified two galaxies with statistically significant X-ray detections: UHZ\,1 ($\rm{RA}=3.567067$, $\rm{Dec}=-30.377869$) at a spectroscopic redshift of $z_{\mathrm{spec}}=10.07$ \citep{bogdan2023detection,Goulding+2023}, and GHZ\,9 ($\rm{RA} = 3.478739$, $\rm{Dec}=-30.345535$) at a photometric redshift of $z_{\mathrm{phot}}=10.37^{+0.32}_{-1.09}$ \citep{atek23}. While UHZ\,1 is detected in the hard band, GHZ\,9 is primarily detected in the soft band. Our previous paper describes the X-ray detection and inferred properties of a heavily obscured active galactic nucleus (AGN) in UHZ\,1 \citep{bogdan2023detection}. Here, we focus on GHZ\,9 (see Section \ref{sec:ghz9} for its detailed properties), and present the candidate X-ray AGN associated with it. Other, less significant detections and non-detections of the $z>9$ galaxies behind Abell\,2744, will be discussed in a follow-up paper.

We assume the following values for the standard cosmological parameters: $H_0=71 \ \rm{km \ s^{-1} \ Mpc^{-1}}$, $ \Omega_M=0.27$, and $\Omega_{\Lambda}=0.73$, and all error bars represent $1\sigma$ uncertainties.

\section{The GHZ\,9 Galaxy}
\label{sec:ghz9}

GHZ\,9 (also known as galaxy ID 52008 in Atek et al. 2023) is a Lyman-break galaxy with a dropout in the F115W NIRCam bandpass (Figure\,\ref{fig:ghz9_jwst}), which suggests its high-redshift nature.
Its photometric redshift, stellar mass and star formation rate are obtained by spectral energy distribution (SED) modeling.

The templates used in SED modeling do not include the contribution from AGN or broad emission lines; the former is highly uncertain even at low redshift and line emission can be very different at the high ionization rates inferred in these highly star-forming galaxies. With these caveats in mind, we note that the redshift of GHZ\,9 ranges from $10.37^{+0.32}_{-1.09}$ (\citet{atek23}, \textsc{eazy} code) through $9.47^{+0.36}_{-0.35}$ (\citet{atek23}, \textsc{beagle} code) to as low as $9.35^{+0.77}_{-0.35}$ (\citet{castellano22b}, \textsc{eazy} code).
Note that these redshifts are in agreement within $1\sigma$ uncertainty.
Similarly, the stellar mass ranges from $ (4.9^{+2.0}_{-2.6})\times 10^7\, \mathrm{M_{\odot}}$ \citep{atek23} to $(3.3^{+2.9}_{-2.4})\times 10^8 \, \mathrm{M_{\odot}}$ \citep{castellano22b}, and the star formation rate from $(0.56^{+0.23}_{-0.29}) \, \mathrm{M}_{\odot} \, \mathrm{yr}^{-1}$ \citep{atek23} to $(14.4^{+15.0}_{-7.3}) \, \mathrm{M}_{\odot} \, \mathrm{yr}^{-1}$ \citep{castellano22b}. 

A potential concern is that the probability distribution function (PDF) for the photometric redshift of GHZ\,9 shows a secondary peak at $z \approx 2$.
Follow-up measurements with NIRCam will conclusively determine the redshift. We note that systematic issues, such as the template choices mentioned above, would only act to broaden the PDF around the primary peak. For the rest of this paper, we proceed under the assumption that the true redshift of GHZ\,9 is $\sim10.4$ from \citet{atek23}.

\citet{2024ApJS..270...12W} performed SED modeling of all \textit{JWST}-detected galaxies in the Abell~2744 field using the \textsc{prospector} package. In their model, they also included a potential AGN contribution, which is further discussed in Section\,\ref{sec:fagn}. For GHZ\,9, they obtained a best-fit redshift of $11.07^{+0.25}_{-0.26}$, a stellar mass of $(2.77^{+1.7}_{-0.9})\times 10^8\, \mathrm{M_{\odot}}$, and a star formation rate of $(5.81^{+1.10}_{-0.99}) \, \mathrm{M}_{\odot} \, \mathrm{yr}^{-1}$. These properties are consistent with those obtained without an AGN component.

The lensing magnification at the position of GHZ\,9 is $\mu=1.26^{+0.02}_{-0.02}$ \citep{Atek_2023}, which we use to correct the observed quantities.

\section{Analysis of the \textit{Chandra}  data}
\label{sec:data}

We utilized 98 \textit{Chandra} ACIS observations centered on the core of the galaxy cluster Abell\,2744. The total available exposure time of the data is $2.09$~Ms (Table \ref{tab:data}). The analysis was carried out using standard \textsc{CIAO} tools (version 4.15) and the CALDB version 4.10.7 \citep{2006SPIE.6270E..1VF}.

The main steps of the data reduction were identical to those presented in \citet{bogdan22}. First, we reprocessed the individual observations using the \textsc{chandra\_repro} tool. We filtered high-background periods, which resulted in a $3.3\%$ reduction in the total exposure time.
To correct for minor astrometric variations across the individual observations and to minimize the aspect difference between individual data sets, we applied the \textsc{wcs\_match} tool. Finally, the observations were merged using the \textsc{merge\_obs} task, which also accounts for the various roll angles of individual observations.

We probed the accuracy of absolute astrometry by comparing the centroids of 25 \textit{Chandra}--\textit{JWST} source pairs with off-axis angles varying in a wide range from sources close to the aim point to sources with  $\sim7\arcmin$ offset. The X-ray sources have a median detection significance of $\sim8\sigma$ on the merged $0.5-3$\,keV band \textit{Chandra} image, while the \textit{JWST} sources in these pairs have a median brightness of $\sim 68\,\rm{nJy}$ in the F200W band. We found an average offset of $\delta = 0.26\arcsec$ with a dispersion of $\rm{rms} = 0.17$. The largest individual offset among these sources was $0.66\arcsec$. These results agree with those obtained in previous studies \citep[e.g.][]{2021A&A...648A..47L}.

We generated exposure maps for each observation assuming a power-law model with a slope of $\Gamma=2.3$, which is typical for high-redshift AGN \citep{vito19,wang21}, and with a Galactic column density of $N_{\rm{H}} = 1.6 \times 10^{20} \ \rm{cm^{-2}}$. 
The \textit{Chandra} point spread function (PSF) varies across the field-of-view, and GHZ\,9 is located at $\sim6\arcmin$ offset from the aim point at all observations, implying a broad PSF. 
To construct the PSF for each observation, we ran the MARX\footnote{https://space.mit.edu/cxc/marx/} ray-trace simulator through the \textsc{simulate\_psf} \textsc{CIAO} tool for a monochromatic source with $1.5$\,keV energy for each observation. In the $0.5-3$~keV band, the mean PSF radius with $90\%$ enclosed counts fraction is $\sim\!4.5\arcsec$ at the location of GHZ\,9.

\begin{figure}
\centering
\includegraphics[width=.49\textwidth]{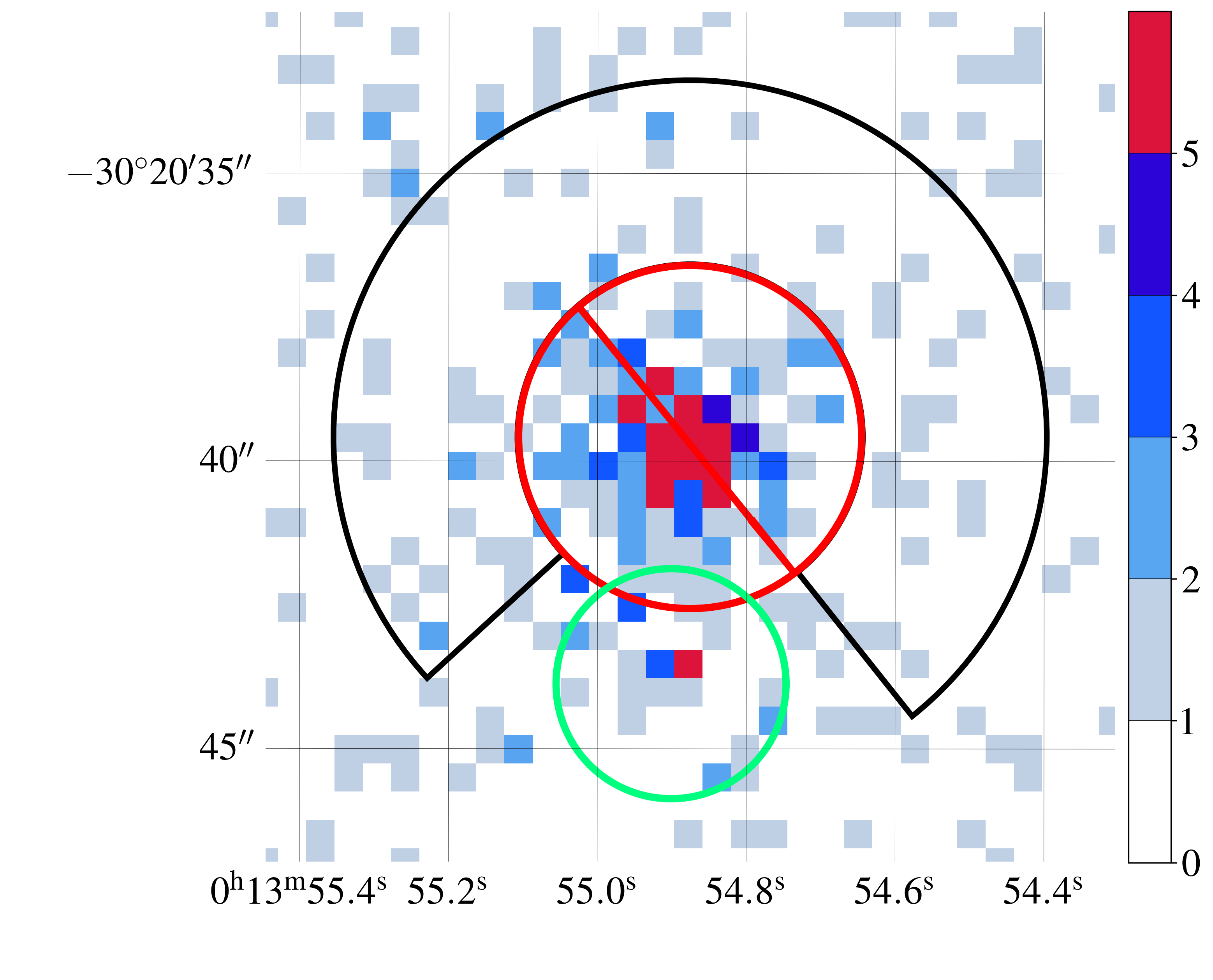}
\caption{$0.5-3$\,keV band \textit{Chandra} image of GHZ\,9-X and its surroundings centered around the neighboring star. A source region of a $r=2\arcsec$ circle (green region) and a background wedge centered around the star (black region) were applied to account for the net counts in our second approach (Section\,\ref{sec:detection}). A circle with $r=3\arcsec$ around the star (red region) was excluded from the analysis. The colorbar indicates the scaled number of counts.}
\label{fig:ghz9_chandra}
\end{figure}

\section{Results}
\label{sec:results}

\subsection{An X-ray source associated with GHZ9}
\label{sec:detection}

In Figure \ref{fig:ghz9_jwst}, we present the $0.5-3$~keV band \textit{Chandra} X-ray and \textit{JWST} near-infrared images of GHZ\,9 and its surroundings. An X-ray source co-spatial with GHZ\,9 is detected in the $0.5-3$\,keV band but remains undetected in the $3-7$\,keV band. The centroid of the X-ray source is $\approx0.24\arcsec$ from the center of GHZ\,9, which offset is consistent with the accuracy of the astrometry (Section \ref{sec:data}).

Panel (c) of Figure \ref{fig:ghz9_jwst} also shows the presence of a bright star \citep[Gaia ID 2320218780150294272][]{2013yCat.1324....0S} located $\approx\!4.4\arcsec$ toward the north from GHZ\,9. The star is detected both in the \textit{Chandra} and \textit{JWST} images. Although the star is cleanly separated from GHZ\,9 in the \textit{JWST} images, the broad \textit{Chandra} PSF (Section\,\ref{sec:data}) introduces complexity to the X-ray analysis as the bright star is expected to contribute to the observed emission at the position of GHZ\,9\footnote{At the location of the bright star, the simulated PSF has a slightly elliptical shape, elongated in the northeast-southwest direction, thus the major axis is not aligned with GHZ\,9-X.}. Finally, we also highlight another low-redshift galaxy about $0.55\arcsec$ to the southwest from GHZ\,9, which galaxy is further discussed in Section \ref{sec:other_gxy}.
For clarity, we refer to the X-ray point source co-spatial with GHZ\,9 as GHZ\,9-X, and to the southwest galaxy as SW galaxy throughout the rest of the paper.

To derive the brightness of GHZ\,9-X, we carried out forced photometry at the position of GHZ\,9. To this end, we extracted the counts from the soft and hard band \textit{Chandra} images using a circular region that is centered on GHZ\,9. We also exclude an $r=3\arcsec$ circle centered at the star (see Figure\,\ref{fig:ghz9_chandra}). Within this source region, we measured $26$ and $8$ counts in the $0.5-3$\,keV and $3-7$\,keV bands, respectively. To account for the background emission and derive the number of net counts associated with the source, we employ two approaches.

In the first approach, we describe the background with two components: (i) the background emission originating from the ICM of Abell\,2744, and the sky\footnote{Sky background includes the cosmic X-ray background, and the diffuse emission from the Milky Way and the Local Hot Bubble.} and instrumental background components, (ii) spillover counts originating from the nearby star.
Note that GHZ\,9-X is located at a $\sim 6 \arcmin$ projected distance from the cluster core, where the emission from the intracluster medium (ICM) drops to $\sim1\%$ of the core value. 
At this radius, the instrumental background dominates the overall background \citep{2023A&A...678A..91K}, and surpasses both the ICM emission and the sky background.
To account for the background emission, we used a local, $40\arcsec \times 40\arcsec$, rectangular region around GHZ\,9-X, excluding $r=5.5\arcsec$ and $r=3\arcsec$ circular regions around the star and GHZ\,9-X, respectively. We assigned $r=2\arcsec$ circular regions centered on each pixel within the rectangle, which resulted in 328 background regions. We extracted the number of counts within each region and found a mean value of $5.5$ counts after scaling the areas. We emphasize that none of the individual circular background regions had $26$ counts.
To derive the number of spillover counts from the star, we rely on the simulated \textit{Chandra} PSF normalized to the stellar brightness. Specifically, we extracted the $0.5-3$~keV band radial surface brightness profile of the star using circular annuli. Then, we matched the distribution of the stellar profile with that obtained from the simulated PSF. We obtained excellent agreement between the observed and simulated profiles, confirming that the star is a point source. 
Because the extraction region of GHZ\,9-X is located at $3\arcsec-6.1\arcsec$ offset from the star (Figure\,\ref{fig:ghz9_chandra}), we integrated the normalized simulated PSF distribution within this annulus, and scaled the result to the area of the GHZ\,9-X region. We concluded that the star contributes with $\sim\!4.8$ counts. 
Thus, the spillover counts from the star and the background emission contribute a total of $10.3$ counts to the GHZ\,9-X extraction region. Given the $26$ total counts in the source region, we conclude that GHZ\,9-X has $15.7$ net counts. Assuming Poisson distribution, the probability of detecting $26$ counts, given the above estimated background level, is $1.8\times10^{-5}$, which corresponds to a detection significance of $4.3\,\sigma$.
In the $3-7$\,keV band, the X-ray source is not detected with a high statistical significance, hence we derive a $1\sigma$ upper limit of $<4.4$ counts. This non-detection is consistent with the assumed power law model (Section \ref{sec:data}).

In the second approach, we estimated the two background components simultaneously from the star's PSF wing with $3\arcsec-6.1\arcsec$ radii (Figure \ref{fig:ghz9_chandra}) measured in the observed image. Within this wedge, we defined 70 overlapping circular regions with radius and offset identical to that of the GHZ\,9-X source region, and extracted the counts within each circle. The counts associated with the star within a $3\arcsec$ region were excluded from this analysis. The mean $0.5-3$\,keV background level within the wedge is $7.5$ counts and no regions have $26$ counts. We thus conclude that $18.5$ net counts are associated with GHZ\,9-X in the soft band, which provides a detection probability of $7.5 \times 10^{-8}$ and a detection significance of $5.4\sigma$, assuming Poisson distribution. In agreement with the previous approach, the source is not significantly detected in the $3-7$~keV band.

Both approaches reveal comparable detection significances ($4.3\sigma$ and $5.4\sigma$) in the soft band, and indicate the presence of an X-ray source co-spatial with GHZ\,9.

\subsection{Hardness ratios}
\label{sec:spectrum}

To further test whether GHZ\,9-X is distinct from the nearby star, we compare the spectral characteristics of the two sources.
Due to the relatively low number of counts originating from GHZ\,9-X, we limit our analysis to deriving X-ray hardness ratios. We define the hardness ratio as $H\!R=\frac{A-B}{A+B}$, where $A$ and $B$ are the number of counts measured in the $1.5-3 \,\rm keV$ and $0.5-1.5 \,\rm keV$ bands, respectively. We calculated the hardness ratios and the associated confidence limits using the \texttt{BEHR} (Bayesian Estimation of Hardness Ratios) code \citep{2006ApJ...652..610P} for GHZ\,9-X and the star using the regions defined in Section\,\ref{sec:detection} and shown in Figure\,\ref{fig:ghz9_chandra}. For GHZ\,9-X, the background counts were extracted from the annular wedge (Figure\,\ref{fig:ghz9_chandra}), while for the star we used a nearby circular region with $r=10\arcsec$ radius.

We obtained hardness ratios of $\rm{HR}  = 0.23 \pm 0.09$ and $\rm{HR}  = -0.30_{-0.32}^{+0.41}$, for the star and GHZ\,9-X, respectively. This suggests that GHZ\,9-X does not have the same spectral characteristics as the star. The observed hardness ratio for GHZ\,9-X is consistent with an unobscured or mildly obscured power law model with a slope of $\Gamma = 2.3$. However, due to the relatively low number of counts and the uncertainties associated with the hardness ratios, we cannot further characterize the spectral properties of the source. 

\begin{figure}[t]
  \begin{center}
    \leavevmode
      \epsfxsize=0.48\textwidth \epsfbox{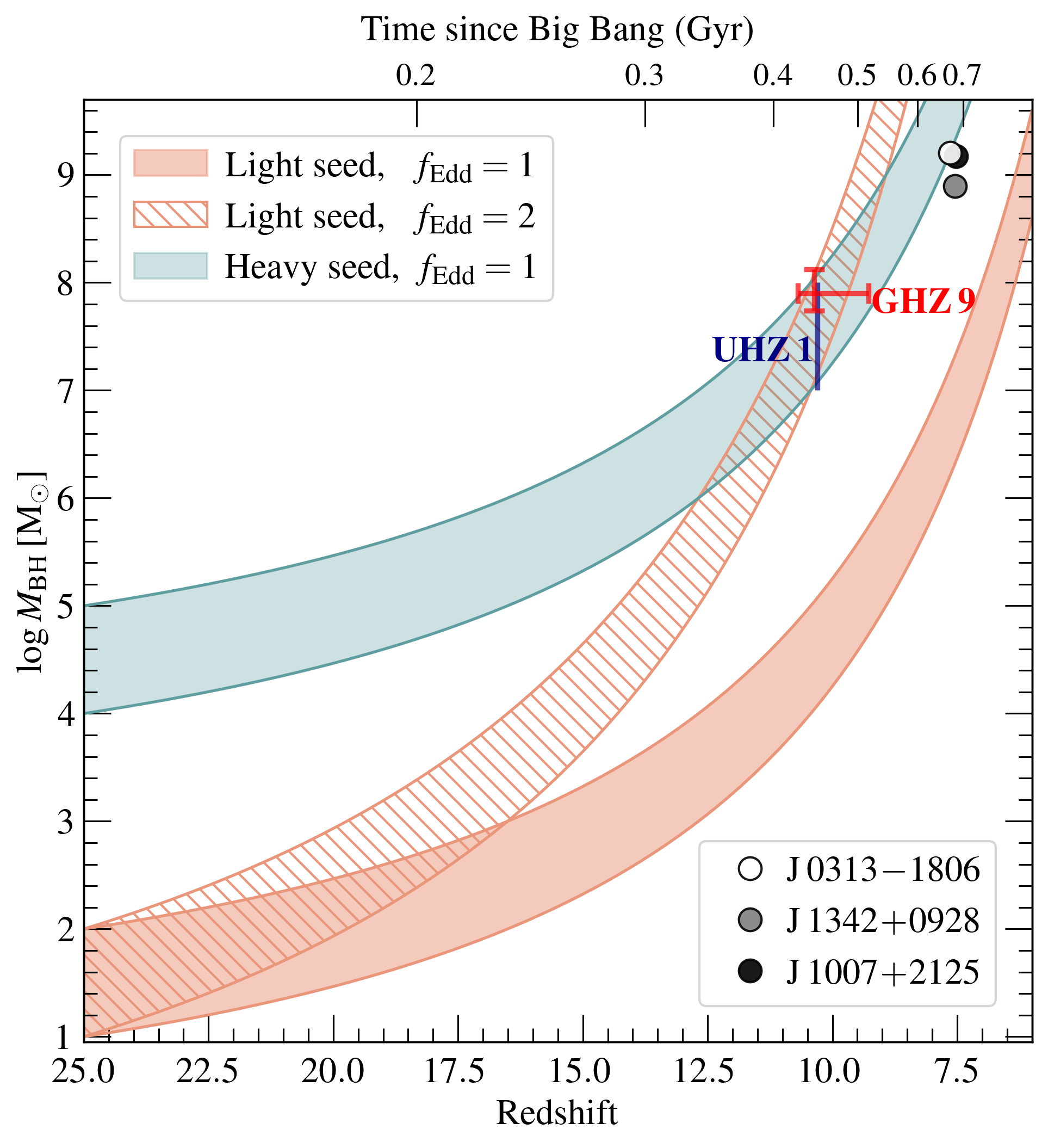}
      \vspace{0cm}
       \caption{BH growth tracks for different initial seed masses and accretion rates assuming continuous accretion with a radiative efficiency of 10\%. Light BH seeds only reach $10^4-10^5\,\rm M_{\odot}$ mass by $z\approx10.4$ accreting at their Eddington limit. The estimated $\sim\!8.0\times10^{7}\,\rm M_{\odot}$ mass for its redshift of $z\approx10.37$, however, places the candidate BH of GHZ\,9 as originating from a heavy seed, similar to UHZ\,1\,\citep{bogdan2023detection}. While sustained accretion above the Eddington limit is unlikely \citep{2010AJ....140..546W,Smith+2018}, we also show the growth curves assuming $f_{\mathrm{Edd}}=2$ for light seeds with the hatched region, which would also be able to produce BH masses estimated for GHZ\,9 and UHZ\,1. We show three previously known highest redshift quasars at $z\sim\!7.5$ identified in large-area optical surveys\,\citep{2011Natur.474..616M,2018Natur.553..473B,2021ApJ...907L...1W}.}
     \label{fig:growth_curve}
  \end{center}
\end{figure}

\subsection{The luminosity and BH mass of the candidate X-ray AGN}
\label{sec:brightness}

GHZ\,9-X has $15.7-18.5$ net counts within the $2\arcsec$ extraction radius in the $0.5-3$\,keV band (Section~\ref{sec:detection}). We use the mean of these values and the Gehrels approximation to derive its confidence limit \citep{gehrels86}, which yields $17.1\pm5.2$ net counts. As the spectral characteristics of the galaxy do not indicate high-level obscuration, we assume a Galactic column density of $N_{\rm{H}} = 1.6 \times 10^{20} \ \rm{cm^{-2}}$ \citep{2013MNRAS.431..394W}. Additionally, we consider that the $2\arcsec$ source region encircles $\sim\!60\%$ of the counts due to the local PSF. Finally, we correct for the $\mu=1.26$ lensing magnification.
Thus, we obtain a $0.5-3$\,keV band (restframe $5.7-34.1$\,keV) flux of $(2.4\pm0.7) \times10^{-16} \ \rm{erg \ s^{-1} \ cm^{-2}}$.

To derive the bolometric luminosity of GHZ\,9-X, we utilize the $L_{\rm bol}/L_{\rm 0.5-2keV}$ correction derived by \citet{lusso12}. Therefore, we first compute the $0.5-2$~keV band luminosity based on the  $0.5-3$~keV band flux and the assumed power law model with $\Gamma=2.3$, which results in $L_{\rm 0.5-2keV} = (2.9\pm0.9) \times 10^{44} \ \rm{erg \ s^{-1}}$. Applying the bolometric correction of $L_{\rm bol}/L_{\rm 0.5-2keV} = 34.9^{+3.9}_{-4.5}$ leads to a bolometric luminosity of $L_{\rm bol} = (1.0^{+0.5}_{-0.4})\times10^{46} \ \rm{erg \ s^{-1}}$. We note that converting the soft band fluxes to the $2-10$~keV band and using the $L_{\rm bol}/L_{\rm 2-10keV}$ correction yields similar bolometric luminosities. Assuming accretion at the Eddington limit, the obtained bolometric luminosity corresponds to a BH mass of $M_{\rm BH} = (8.0^{+3.7}_{-3.2})\times10^7 \ \rm{M_{\odot}}$.

The uncertainties quoted for the bolometric luminosity and BH mass are statistical; they do not include systematic uncertainties. Systematic uncertainties could result from the not well-constrained slope of the power-law model: for example, a $\Gamma=1.9$ slope would give a $\sim20\%$ lower X-ray flux. The uncertain redshift of GHZ\,9 could also affect the observed quantities; if the redshift was $z\approx9.4$, the luminosity and mass of the BH would be $\sim20\%$ lower. Additional uncertainties arise from the intrinsic scatter in the $L_{\rm bol}/L_{\rm 0.5-2keV}$ relation \cite{lusso12}, from potential intrinsic absorption that would lead to a higher inferred luminosity and BH mass, and from potential deviations from the assumed Eddington-limited accretion. Specifically, in the case of super-Eddington accretion, the BH mass could be up to a few times smaller, while lower accretion rates and the presence of intrinsic absorption could result in significantly larger BH masses. Follow-up \textit{JWST} spectroscopic observations could substantially reduce these uncertainties and provide a more accurate BH mass measurement.

\subsection{Broadband SED of GHZ\,9}
\label{sec:fagn}
Recently, \citet{wang24} performed photometric SED fitting on over 60,000 galaxies in the Abell~2744 field. They used the \textsc{prospector} code to determine galaxy properties, including a model that accounted for a potential AGN component. This approach provides an estimate of the contribution of the AGN to the observed light by \textit{JWST}. For GHZ\,9,  the AGN fraction was estimated at $f_{\rm AGN} = (0.2^{+14.7}_{-0.2})\%$. We emphasize that this SED fitting was exclusively based on the \textit{JWST} data and did not consider the X-ray detection discussed in this work.

By combining the \textit{Chandra} X-ray and \textit{JWST} near-infrared data, we conducted a preliminary SED fitting employing the accretion disk model described by \citet{kubota18}, following the methodology outlined by \citet{azadi23}. The accretion disk SEDs were primarily constrained by the X-ray data given the high luminosity of GHZ9-X, which suggests that the X-ray emission is unlikely to be dominated by the host galaxy. Indeed, the broadband SED indicates that GHZ9-X is likely a Type-2 AGN, consistent with the findings of \citet{wang24}, confirming that the \textit{JWST} emission predominantly arises from the host galaxy rather than the AGN accretion disk.

Assuming the Milky Way dust extinction curve and assuming that the dust grains in the AGN torus are similar to those in the interstellar medium, we estimate that a column density of $\sim3\times10^{21} \ \rm{\rm cm^{-2}}$ could effectively obscure the AGN signal in the \textit{JWST} bands. 
We emphasize that this level of column density will not significantly affect the X-ray data measured in the restframe $5.7-34.1$~keV band, hence the inferred X-ray luminosity and BH mass presented in Section \ref{sec:brightness} will remain invariant. Our preliminary SED modeling included only the accretion disk component, however, a comprehensive modeling approach requires both the accretion disk and host galaxy components, with corrections made for the torus and host galaxy absorption affecting the visible/UV photometry associated with the accretion disk. Although full SED modeling of GHZ\,9 is beyond the scope of the present work, our conclusions are well constrained by the preliminary SED modeling.

\subsection{Random galaxy simulations}
\label{sec:simulation}

Because we searched for X-ray sources co-spatial with 19 \textit{JWST}-identified galaxies, we assess the likelihood that GHZ\,9-X is a random upwards fluctuation. First, we employ the \textsc{CIAO} \textsc{wavdetect} tool to determine the detection significance of the X-ray source in the $0.5-3$~keV band. Because the vicinity of GHZ\,9 has a low-level emission from the galaxy cluster, we only ran \textsc{wavdetect} in a $50\arcsec\times50\arcsec$ region around the galaxy. The X-ray source was detected at a significance threshold of $7\times10^{-5}$. Projecting this result to the entire ACIS-I array, we expect $\approx290$ false detections.

Next, we assess the probability distribution of getting a chance alignment between a \textit{JWST} source and an X-ray source in the \textit{Chandra} data. To this end, we simulate $10,000$ random X-ray sources and use the full catalog of $61,648$ \textit{JWST} sources in the Abell\,2744 field \citep{2023arXiv230102671W} to measure the distribution of separation distances between the simulated X-ray sources and the \textit{JWST} sources. Because the offset between the X-ray source and GHZ\,9 is $\approx0.24\arcsec$, we integrate the obtained distribution out to $<0.24\arcsec$. Then, we normalize this value by the number of simulated random X-ray sources and the expected number of false detections, take into account the ratio between the \textit{JWST} and \textit{Chandra} field-of-views, and account for the fact that we searched for X-ray sources at $19$ $z>9$ \textit{JWST} galaxies. We thus find that the likelihood of having a falsely detected X-ray source within $0.24\arcsec$ from the position of a high-redshift \textit{JWST} galaxy is $4.7\times10^{-4}$, corresponding to $\sim\,3.5\sigma$.

\subsection{A low-redshift galaxy in the proximity of GHZ\,9}
\label{sec:other_gxy}

Situated at $0.55\,\arcsec$ angular distance to the southwest from GHZ\,9, a  galaxy \citep[][]{2023arXiv230102671W} is resolved in the \textit{JWST} images (Figure\,\ref{fig:ghz9_jwst}). The same galaxy is detected in the Hubble Frontier Field Deep Space catalog \citep{2018ApJS..235...14S,2021MNRAS.506..928N} and described as a low-stellar-mass galaxy ($\sim\!10^8 \rm M_{\sun}$) with a photometric redshift of $z\sim 1.6$. We discuss if GHZ\,9-X could be associated with this lower redshift galaxy.

Based on the distribution of the $0.5-3$~keV band counts around GHZ\,9-X, we find that its best-fit centroid is $\rm{RA}=3.4787792$ and $\rm{Dec}=-30.345478$. This corresponds to projected distances of $\sim0.24\arcsec$ from GHZ\,9 and $\sim0.8\arcsec$ from the SW galaxy. This suggests that GHZ\,9-X is co-spatial with GHZ\,9. The accuracy of the astrometry, and hence the association between GHZ\,9-X and GHZ\,9, is further supported by the relative distances between the bright star and GHZ\,9-X on the \textit{Chandra} image and between the bright star and GHZ\,9 on the \textit{JWST} image. The importance of this is that even though there may be differences in the absolute astrometry between the two space telescopes, the relative distances should be identical. On the \textit{Chandra} image, the projected distance between the centroid of the bright star and GHZ\,9-X is $\approx3.9\arcsec$, while on the \textit{JWST} image, the projected distance between the centroid of the star and GHZ\,9 is $\approx4.2\arcsec$. This small difference (about half a \textit{Chandra} pixel) between the relative distances suggests that the X-ray source is associated with GHZ\,9 rather than the SW galaxy. Indeed, the projected distance between the bright star and the SW galaxy is $\approx4.7\arcsec$ on the \textit{JWST} image, which is substantially larger than that measured between the bright star and GHZ\,9-X on the \textit{Chandra} image. 

As a caveat, we note that systematic studies of deep \textit{Chandra} fields, which explored larger samples of X-ray-optical source pairs, suggest that the positional uncertainty depends on the \textit{Chandra} off-axis position and the brightness of the source  \citep[e.g.][]{2005ApJS..161...21L,2012ApJS..202....6G}. Based on these works and considering that GHZ\,9-X resides $\sim6\arcmin$ from the ACIS-I aim point and is a relatively faint source, the systematic positional uncertainty could be as large as $\approx1\arcsec$. We note that if GHZ\,9-X were to be associated with the SW galaxy, its X-ray luminosity would be $L_{\rm 0.5-2keV} = (4.1\pm1.3) \times 10^{42} \ \rm{erg \ s^{-1}}$.

Taken together, we conclude that while the X-ray source is most likely associated with GHZ\,9, due to potential systematic uncertainties in the astrometry, we cannot completely rule out that the X-ray source is associated with the SW galaxy.
Follow-up \textit{JWST} spectroscopic observations could conclusively determine if GHZ\,9 hosts an accreting BH.

\section{Discussion \& Conclusions}
\label{sec:discussion}
Considering the evidence presented in Section \ref{sec:results}, we assume that GHZ\,9-X is a rapidly growing BH. If confirmed by follow-up observations, at $z\sim10.4$ (i.e.\ merely $\sim450$ Myrs after the Big Bang), it would be one of the earliest accreting BHs ever detected. In this section, we discuss the potential origin and the theoretical implications of the BH candidate. 

To infer the origin of this BH, we track the assembly history of light and heavy seeds (Figure  \ref{fig:growth_curve}). A light initial seed accreting at its Eddington rate for 330~Myrs (from $z=25$ to $z=10.4$), would only reach a final mass of $\sim10^{4-5} \ \rm{M_{\odot}}$, falling short of the inferred $8\times10^7 \ \rm{M_{\odot}}$ BH mass by $2-3$ orders of magnitude. To reach this mass, a light seed must continuously accrete at least twice the Eddington rate, which is not supported by simulations \citep[see the discussion below and][]{2010AJ....140..546W,Smith+2018}. In contrast, heavy seeds accreting at or somewhat below the Eddington rate can easily reach the BH mass of GHZ\,9-X within the available cosmic timeframe.

\begin{figure}[]
  \begin{center}
    \leavevmode
      \epsfxsize=.5\textwidth \epsfbox{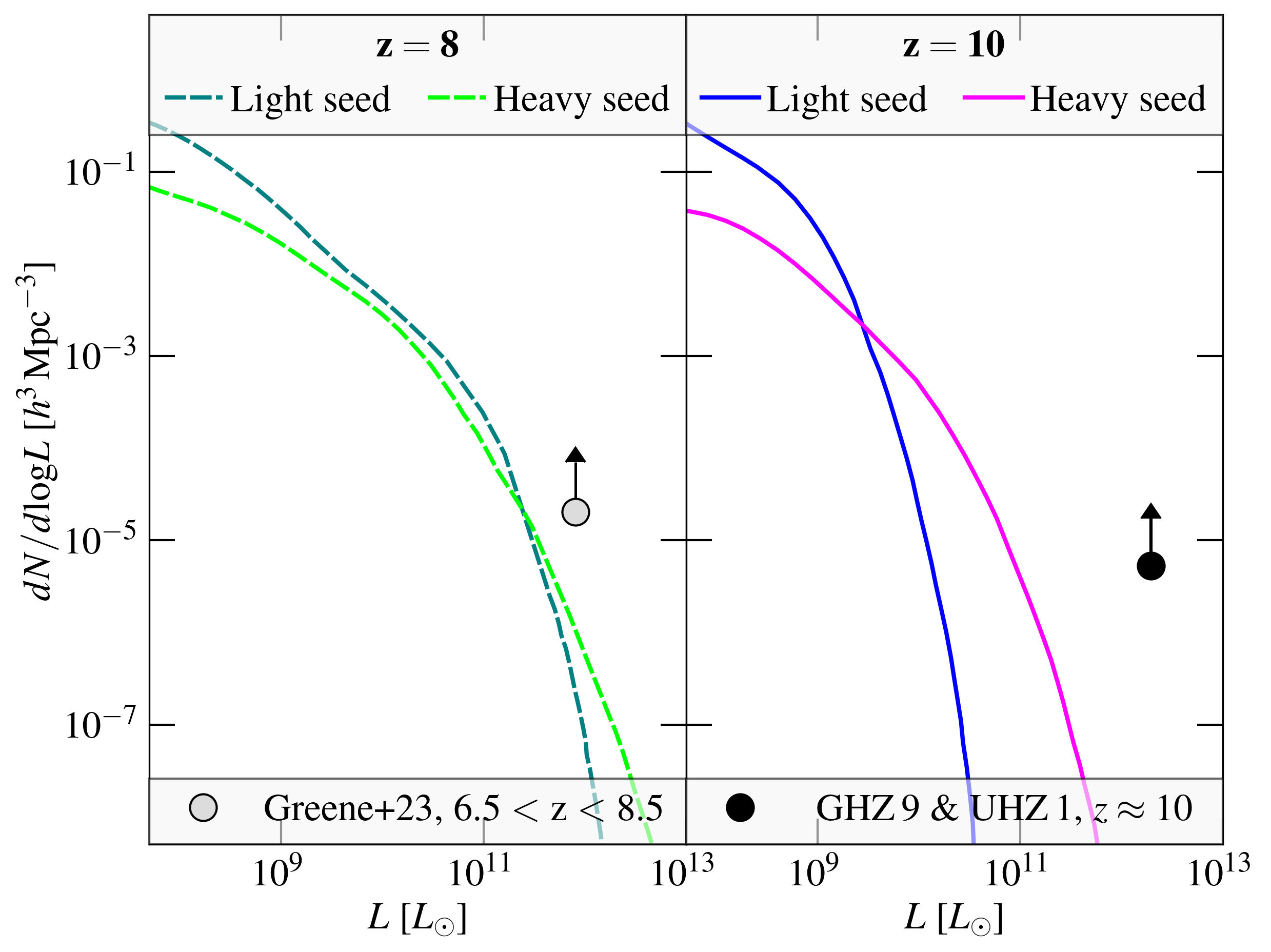}
      \vspace{0cm}
      \caption{Predicted luminosity function of AGN at $z=8$ (left panel) and $z=10$ (right) for light and heavy BH seed models \citet[adapted from][]{Ricarte+2018}. Overplotted are the observed lower limits on the number density at $z=8$ \citep{Greene+2023} and at $z=10$ based on the detections of AGN behind Abell\,2744 (i.e.\ UHZ\,1 and GHZ\,9). The number density of AGN at $z=10$ drastically exceeds theoretical expectations.} 
     \label{fig:lf}
  \end{center}
\end{figure}

While there are likely multiple mechanisms available for forming direct collapse BHs, the halos where they form are expected to be rare peaks and clustered at early times. Therefore, the detection of two Outsize Black Hole Galaxy (OBG) candidates (GHZ\,9 and UHZ\,1) behind the same lensing cluster is unsurprising. Taken together, these two detections suggest that the formation of direct collapse BHs might be more efficient than previously assumed, adding support to the existence of multiple potential formation pathways, such as cold gas flows in early halos \citep{Latif+2022}, mergers of massive gas-rich galaxies in the early Universe \citep{Mayer+2023}, or early star-burst clusters \citep{kroupa2020}. In more exotic scenarios, some models predict the emergence of primordial BHs with masses of $10^6 \,\textrm{M}_{\odot}$ \citep[e.g.][]{carr2021,escriva2022}.

Figure \ref{fig:growth_curve} shows that the light seed channel is only feasible with sustained super-Eddington accretion for hundreds of Myrs (Figure \ref{fig:growth_curve}). Although state-of-the-art simulations cannot track the formation and evolution of these seeds ab initio, they offer insights into the growth of early BHs. Using the Renaissance simulation suite, \citet{Smith+2018} followed the growth of over 15,000 light seeds and found that these seeds grow extremely inefficiently, with the most active BHs growing by $10\%$ over 300~Myrs. Of this large ensemble, only one seed experienced a brief period of super-Eddington accretion. Thus, \citet{Smith+2018} concluded that light seeds cannot reach  $\sim10^7-10^8 \ \rm{M_{\odot}}$ by $z\sim 10$. A similar conclusion was found based on the CAT (Cosmic Archaeology Tool) semi-analytic model, which found that light BH seeds cannot reach the masses and luminosities of quasars observed at $z\gtrsim6$  \citep{2022MNRAS.511..616T}.

The \textit{Chandra}/\textit{JWST} discovery of $z\sim8-10$ AGN paves the way for building the luminosity function of AGN in the early universe and comparing the observational data with theoretical predictions. Current theoretical studies utilize semi-empirical models/semi-analytic models to trace the formation and assembly history of BH populations over cosmic time. These works vary both the BH seeding and the growth prescriptions to discern potential surviving observational signatures of initial BH seeding. Notably, \citet{Ricarte+2018} predict bolometric luminosity functions for accreting BHs at a range of redshifts for the light and heavy seed models.
In Figure \ref{fig:lf}, we compare the \citet{Ricarte+2018} model predictions of AGN luminosity functions at $z\sim10$ based on UHZ\,1 and GHZ\,9 and at $z= 6 - 8.5$ using data from \citet{Greene+2023}. The observed AGN number density at $z=6 - 8.5$ exceeds the theoretical expectations for the light and heavy seed models. We note that the AGN luminosity functions at $z\sim8$ are nearly identical, hence AGN at these redshifts cannot be used to discriminate between seeding models. The difference relative to the model predictions becomes striking at $z=10$, where the observed AGN number density exceeds the theoretically predicted values by several orders of magnitude. This suggests that the efficiency of heavy seed formation is theoretically underestimated, hinting that multiple channels for heavy seed formation may simultaneously operate in the early universe. We note that the presented luminosity function is based on two $z\sim10$ sources that were detected in the single, and relatively small, field of Abell 2744. To establish a more accurate luminosity function of $z\sim10$ AGN, the sample size of high-redshift AGN should be increased, and cosmic variance should be explored by adding other sightlines.

\bigskip

\begin{small}
\noindent
\textit{Acknowledgements.}
We thank the reviewer for their constructive report. O.E.K and N.W. are supported by the GA\v{C}R EXPRO grant No. 21-13491X. \'A.B., M.A., W.R.F, C.J., and R.P.K acknowledge support from the Smithsonian Institution and the Chandra Project through NASA contract NAS8-03060. PN acknowledges support from the Black Hole Initiative at Harvard University, which is funded by grants from the John Templeton Foundation and the Gordon and Betty Moore Foundation.
This paper employs a list of \textit{Chandra} datasets, obtained by the \textit{Chandra} X-ray Observatory, contained in~\dataset[DOI: X]{https://doi.org/10.25574/cdc.194}.
\textit{JWST} data presented in this article were obtained from the Mikulski Archive for Space Telescopes (MAST) at the Space Telescope Science Institute, which can be accessed via \dataset[DOI]{https://dx.doi.org/10.17909/kw3c-n857}.
\end{small}

\bibliographystyle{aa}
\bibliography{paper1} 

\begin{appendix}
We list the analyzed \textit{Chandra} ACIS observations of Abell\,2744 in Table \ref{tab:data}.

\begin{table}[!t]
\begin{center}
\caption{Analyzed \textit{Chandra} observations of Abell\,2744.}
\begin{minipage}{18cm}
\renewcommand{\arraystretch}{1.3}
\centering
\begin{tabular}{ cccc|cccc}
\hline
Observation ID & $t_{\rm exp}^{\rm clean}\ \rm{(ks)}$ & Detector & Observation Date & Observation ID & $t_{\rm exp}^{\rm clean} \ \rm{(ks)}$ & Detector & Observation Date \\
\hline
2212 	 & 17.82 	 & ACIS-S 	 & 2001-09-03 	 &	 25938 	 & 18.66 	 & ACIS-I 	 & 2022-11-26 	 \\
7915 	 & 18.11 	 & ACIS-I 	 & 2006-11-08 	 &	 25963 	 & 37.59 	 & ACIS-I 	 & 2022-11-26 	 \\
8477 	 & 44.63 	 & ACIS-I 	 & 2007-06-10 	 &	 25937 	 & 30.78 	 & ACIS-I 	 & 2022-11-27 	 \\
8557 	 & 27.30 	 & ACIS-I 	 & 2007-06-14 	 &	 25278 	 & 9.78 	 & ACIS-I 	 & 2022-12-02 	 \\
7712 	 & 8.07 	 & ACIS-I 	 & 2007-09-10 	 &	 27575 	 & 19.65 	 & ACIS-I 	 & 2022-12-02 	 \\
26280 	 & 11.39 	 & ACIS-I 	 & 2022-01-18 	 &	 25936 	 & 12.92 	 & ACIS-I 	 & 2023-01-26 	 \\
25912 	 & 15.10 	 & ACIS-I 	 & 2022-04-18 	 &	 27678 	 & 12.42 	 & ACIS-I 	 & 2023-01-27 	 \\
25911 	 & 16.59 	 & ACIS-I 	 & 2022-04-19 	 &	 25939 	 & 14.32 	 & ACIS-I 	 & 2023-01-28 	 \\
25934 	 & 18.96 	 & ACIS-I 	 & 2022-04-21 	 &	 27679 	 & 11.93 	 & ACIS-I 	 & 2023-01-28 	 \\
25931 	 & 14.55 	 & ACIS-I 	 & 2022-04-23 	 &	 27680 	 & 13.21 	 & ACIS-I 	 & 2023-01-28 	 \\
25954 	 & 13.39 	 & ACIS-I 	 & 2022-04-24 	 &	 27681 	 & 9.78 	 & ACIS-I 	 & 2023-01-29 	 \\
25928 	 & 15.36 	 & ACIS-I 	 & 2022-05-03 	 &	 25909 	 & 19.33 	 & ACIS-I 	 & 2023-05-24 	 \\
25942 	 & 15.18 	 & ACIS-I 	 & 2022-05-04 	 &	 27856 	 & 15.88 	 & ACIS-I 	 & 2023-05-25 	 \\
25958 	 & 11.81 	 & ACIS-I 	 & 2022-05-04 	 &	 27857 	 & 12.92 	 & ACIS-I 	 & 2023-05-26 	 \\
25971 	 & 12.62 	 & ACIS-I 	 & 2022-05-04 	 &	 27563 	 & 11.70 	 & ACIS-I 	 & 2023-06-08 	 \\
25932 	 & 14.08 	 & ACIS-I 	 & 2022-05-05 	 &	 25941 	 & 32.65 	 & ACIS-I 	 & 2023-06-09 	 \\
25972 	 & 31.14 	 & ACIS-I 	 & 2022-05-18 	 &	 27896 	 & 13.73 	 & ACIS-I 	 & 2023-06-10 	 \\
25970 	 & 23.99 	 & ACIS-I 	 & 2022-06-12 	 &	 25917 	 & 35.62 	 & ACIS-I 	 & 2023-06-22 	 \\
25919 	 & 25.03 	 & ACIS-I 	 & 2022-06-13 	 &	 25950 	 & 29.69 	 & ACIS-I 	 & 2023-06-30 	 \\
25920 	 & 29.67 	 & ACIS-I 	 & 2022-06-13 	 &	 25946 	 & 29.69 	 & ACIS-I 	 & 2023-07-01 	 \\
25922 	 & 31.11 	 & ACIS-I 	 & 2022-06-14 	 &	 25965 	 & 35.63 	 & ACIS-I 	 & 2023-07-07 	 \\
25968 	 & 26.92 	 & ACIS-I 	 & 2022-07-12 	 &	 25960 	 & 24.76 	 & ACIS-I 	 & 2023-07-08 	 \\
25967 	 & 33.12 	 & ACIS-I 	 & 2022-08-01 	 &	 25926 	 & 61.18 	 & ACIS-I 	 & 2023-07-12 	 \\
25929 	 & 26.42 	 & ACIS-I 	 & 2022-08-26 	 &	 25955 	 & 43.42 	 & ACIS-I 	 & 2023-07-20 	 \\
25925 	 & 23.41 	 & ACIS-I 	 & 2022-09-02 	 &	 25921 	 & 16.87 	 & ACIS-I 	 & 2023-08-04 	 \\
25956 	 & 13.90 	 & ACIS-I 	 & 2022-09-02 	 &	 25959 	 & 15.39 	 & ACIS-I 	 & 2023-08-05 	 \\
25913 	 & 19.64 	 & ACIS-I 	 & 2022-09-03 	 &	 27974 	 & 28.71 	 & ACIS-I 	 & 2023-08-05 	 \\
25915 	 & 20.31 	 & ACIS-I 	 & 2022-09-03 	 &	 25940 	 & 27.72 	 & ACIS-I 	 & 2023-08-10 	 \\
25923 	 & 10.64 	 & ACIS-I 	 & 2022-09-04 	 &	 25966 	 & 18.84 	 & ACIS-I 	 & 2023-08-13 	 \\
25279 	 & 23.69 	 & ACIS-I 	 & 2022-09-06 	 &	 28370 	 & 20.73 	 & ACIS-I 	 & 2023-08-13 	 \\
25924 	 & 21.54 	 & ACIS-I 	 & 2022-09-07 	 &	 25933 	 & 23.88 	 & ACIS-I 	 & 2023-08-15 	 \\
25944 	 & 20.85 	 & ACIS-I 	 & 2022-09-08 	 &	 28483 	 & 20.22 	 & ACIS-I 	 & 2023-08-19 	 \\
25957 	 & 21.29 	 & ACIS-I 	 & 2022-09-08 	 &	 25935 	 & 24.08 	 & ACIS-I 	 & 2023-08-20 	 \\
27347 	 & 21.45 	 & ACIS-I 	 & 2022-09-09 	 &	 27780 	 & 14.90 	 & ACIS-I 	 & 2023-08-21 	 \\
25918 	 & 20.38 	 & ACIS-I 	 & 2022-09-13 	 &	 25943 	 & 16.69 	 & ACIS-I 	 & 2023-08-31 	 \\
25953 	 & 24.24 	 & ACIS-I 	 & 2022-09-17 	 &	 28872 	 & 13.09 	 & ACIS-I 	 & 2023-09-01 	 \\
25908 	 & 21.83 	 & ACIS-I 	 & 2022-09-23 	 &	 25916 	 & 22.20 	 & ACIS-I 	 & 2023-09-03 	 \\
27449 	 & 9.78 	 & ACIS-I 	 & 2022-09-24 	 &	 25964 	 & 20.32 	 & ACIS-I 	 & 2023-09-05 	 \\
25910 	 & 19.31 	 & ACIS-I 	 & 2022-09-25 	 &	 25961 	 & 18.84 	 & ACIS-I 	 & 2023-09-09 	 \\
27450 	 & 9.78 	 & ACIS-I 	 & 2022-09-26 	 &	 28886 	 & 9.96 	 & ACIS-I 	 & 2023-09-10 	 \\
25945 	 & 16.77 	 & ACIS-I 	 & 2022-09-27 	 &	 28887 	 & 19.85 	 & ACIS-I 	 & 2023-09-10 	 \\
25948 	 & 26.80 	 & ACIS-I 	 & 2022-09-30 	 &	 25962 	 & 21.81 	 & ACIS-I 	 & 2023-09-11 	 \\
25969 	 & 26.92 	 & ACIS-I 	 & 2022-10-09 	 &	 25927 	 & 20.53 	 & ACIS-I 	 & 2023-09-16 	 \\
25914 	 & 27.77 	 & ACIS-I 	 & 2022-10-15 	 &	 25947 	 & 14.90 	 & ACIS-I 	 & 2023-09-24 	 \\
25907 	 & 36.80 	 & ACIS-I 	 & 2022-11-08 	 &	 28920 	 & 15.28 	 & ACIS-I 	 & 2023-09-25 	 \\
25973 	 & 18.15 	 & ACIS-I 	 & 2022-11-11 	 &	 25952 	 & 10.84 	 & ACIS-I 	 & 2023-09-27 	 \\
25930 	 & 19.16 	 & ACIS-I 	 & 2022-11-15 	 &	 28934 	 & 19.83 	 & ACIS-I 	 & 2023-09-29 	 \\
27556 	 & 24.64 	 & ACIS-I 	 & 2022-11-15 	 &	 27739 	 & 21.31 	 & ACIS-I 	 & 2023-10-01 	 \\
25951 	 & 28.45 	 & ACIS-I 	 & 2022-11-18 	 &	 25277 	 & 18.68 	 & ACIS-I 	 & 2023-10-02 	 \\

 \hline
\label{tab:data}
\end{tabular} 
\end{minipage}
\end{center}
\end{table}

\end{appendix}

\end{document}